\documentclass[aps,pra,amssymb,twocolumn,amsmath,superscriptaddress,showpacs,10pt]{revtex4}

\usepackage{graphicx}
\usepackage{dcolumn}
\usepackage{bm}
\usepackage{color}
\usepackage{epstopdf}
\usepackage{subfigure}

\def\TF{\mathrm{TF}}
\def\Or{\mathrm{O}}
\def\U{\mathcal{U}}
\def\K{\mathcal{T}}
\def\e{\mathrm{e}}

\def\Ai{\mathrm{Ai}}

\begin{document}

\title{Domain wall suppression in trapped mixtures of Bose-Einstein condensates}

\author{Francesco V. Pepe}
\affiliation{Dipartimento di Fisica and MECENAS, Universit\`a di
Bari, I-70126 Bari, Italy} \affiliation{INFN, Sezione di Bari,
I-70126 Bari, Italy}

\author{Paolo Facchi}
\affiliation{Dipartimento di Matematica and MECENAS, Universit\`a
di Bari, I-70125 Bari, Italy} \affiliation{INFN, Sezione di Bari,
I-70126 Bari, Italy}

\author{Giuseppe Florio}
\affiliation{Museo Storico della Fisica e Centro Studi e Ricerche
``Enrico Fermi'', Piazza del Viminale 1, I-00184 Roma, Italy}
\affiliation{Dipartimento di Fisica and MECENAS, Universit\`a di
Bari, I-70126 Bari, Italy} \affiliation{INFN, Sezione di Bari,
I-70126 Bari, Italy}

\author{Saverio Pascazio}
\affiliation{Dipartimento di Fisica and MECENAS, Universit\`a di
Bari, I-70126 Bari, Italy} \affiliation{INFN, Sezione di Bari,
I-70126 Bari, Italy}


\begin{abstract}
The ground state energy of a binary mixture of Bose-Einstein
condensates can be estimated for large atomic samples by making
use of suitably regularized Thomas-Fermi density profiles. By
exploiting a variational method on the trial densities the energy
can be computed by explicitly taking into account the
normalization condition. This yields analytical results and
provides the basis for further improvement of the approximation.
As a case study, we consider a binary mixture of $^{87}$Rb atoms
in two different hyperfine states in a double well potential and
discuss the energy crossing between density profiles with
different numbers of domain walls, as the number of particles and
the inter-species interaction vary.
\end{abstract}


\maketitle

\section{Introduction}

Binary mixtures of Bose-Einstein condensates \cite{alkali} are of
great interest due to their complex dynamical features and their
role in the emergence of macroscopic quantum phenomena
\cite{leggett}. Mixtures are experimentally available and usually
made up of two alkali atomic species
\cite{ketterle,myatt,hall,firenze1,firenze2,papp,tojo}. They
generally display repulsive self-interaction and are confined by
various external potentials. Depending on the inter-species
interaction, two classes of stable configurations are possible:
mixed and separated. The latter are more interesting, since it is
in this case that the observation of phenomena such as symmetry
breaking and macroscopic quantum tunnelling of one species through
the other one \cite{aochui,kasamatsu,trippenbach} is possible.
Evidence of phase separation has been observed in
\cite{papp,tojo}. Many recent articles are devoted to the
investigation of dynamical effects in mixtures, such as vortices
and solitons (see \cite{dynamic1,dynamic2,dynamic3,coreless} for
recent experimental and theoretical studies).

Different approaches are possible in order to study the ground
state of these systems. If the number of particles in the
condensate is very large compared to the number of particles in
the excited states, the fields associated to the two species can
be treated as classical wave functions. This approach leads to the
Gross-Pitaevskij equations \cite{stringari}, which are nonlinear
Schr\"{o}dinger equations obtained by finding stationary points of
the zero-temperature grand-canonical energy of the system. The
Thomas-Fermi (TF) approximation, which consists in this case in
neglecting the kinetic energy of the system, is then usually
applied \cite{stringari}. A great deal of results are obtained in
particular cases, such as confinement by a hard wall trap
\cite{wall}, harmonic or lattice potentials \cite{harmonic} and
axisymmetric traps \cite{axis}, also in the presence of the
gravitational force \cite{riboli}. The problem of the stability of
mixtures has been tackled also with renormalization group
techniques \cite{kolezhuk}.

The presence of the kinetic term in the energy functional of a
binary mixtures leads in particular to the regularization of
possible domain walls, which sharply separate the two species in
the TF ground states \cite{aochui,timm,barankov,mazets,van}. This
is generally related to the problem of minimizing the surface
energy, that is also found in the theory of superconductivity
\cite{fetter/landau}.

In this work we introduce a variational method in order to
approximate the Gross-Pitaevskij  solution in a neighborhood of a
domain wall and estimate the total energy of a mixture. Our
technique explicitly takes into account the normalization of the
condensate wave functions and ensures complete analytical
feasibility.

This article is organized as follows: in Section \ref{sec:bm} we
summarize results obtained in the TF approximation which are
relevant to our analysis; in Section \ref{sec:reg} we introduce
the regularization technique and obtain results regarding the
energy increase with respect to the TF approximation; in Section
\ref{sec:dw} we examine a case study in which macroscopic effects
related to domain wall suppression can be observed and present a
quantitative phase diagram; in Section \ref{sec:conc} we suggest
further possible developments of our technique.

\section{Binary mixtures}
\label{sec:bm}

\subsection{Energy functional}

Let us consider a binary mixture of Bose-Einstein condensates,
confined by the external potentials $V_k(x)$, with $k=1,2$. We
assume that the particles are tightly confined in the transverse
directions, so that the system is quasi one-dimensional
\cite{stringari}. Let $U_{kk}>0$ be the parameters that determine
the self interaction between particles of each species and
$U_{12}=U_{21}>0$ the inter-species interaction parameter. The
ground state of the system is determined by the coupled
Gross-Pitaevskij  equations \cite{stringari,fetter/landau,lieb}
\begin{equation}\label{gpeq}
\left\{-\frac{\hbar^2}{2 m_k} \frac{d^2}{dx^2} +V_k(x) + \sum_j
U_{kj}|\psi_j(x)|^2 -\mu_k \right\} \psi_k(x)=0,\displaystyle
\end{equation}
with $k=1,2$. 
They are the variational equations of the quartic energy
functional
\begin{equation}
\mathcal{E}(\psi_1,\psi_2) = \sum_k \{\mathcal{T}_k(\psi_k) +
\mathcal{V}_k(\psi_k) \}+
\mathcal{U}(\psi_1,\psi_2) 
\label{engcpsi}
\end{equation}
under the constraints
\begin{equation}\label{normal}
\int dx\,\rho_k(x) = N_k,
\end{equation}
where
\begin{eqnarray}
 \mathcal{T}_k(\psi_k) &=& \int
\frac{\hbar^2}{2m_k}\left|\frac{d\psi_k}{dx}(x)\right|^2 dx, \nonumber\\
\mathcal{V}_k(\psi_k) &=& \int V_k(x)|\psi_k(x)|^2 dx,
\nonumber\\
\mathcal{U}(\psi_1,\psi_2) &=& \sum_{j,k}
\mathcal{U}_{jk}(\psi_j,\psi_k)
\nonumber\\
&=& \frac{1}{2} \sum_{j,k} \int U_{jk} |\psi_j(x)|^2|\psi_k(x)|^2
dx. \label{engcpsi1}
\end{eqnarray}
Here $\psi_k(x)$ are the condensate wave functions, whose squared
moduli $\rho_k(x)=|\psi_k(x)|^2$ represent the local densities of
each species, $N_k$ are the numbers of particles making up the
condensates, while  $\mathcal{T}_k$, $\mathcal{V}_k$ and
$\mathcal{U}$ are the kinetic, potential and interaction energy,
respectively. We will assume henceforth that the condensate wave
functions are real, since energy minimization requires their
phases to be constant.

\subsection{Thomas-Fermi approximation}
\label{sec:tf}

The TF approximation \cite{stringari} is in this case equivalent
to neglecting the kinetic terms  in (\ref{gpeq})-(\ref{engcpsi}),
so that
\begin{equation}
\label{eq:ETF} \mathcal{E}_{\TF}(\rho_1,\rho_2) = \sum_k
\mathcal{V}_k(\sqrt{\rho_k}) +
\mathcal{U}(\sqrt{\rho_1},\sqrt{\rho_2}) .
\end{equation}
The value of the adimensional parameter
\begin{equation}\label{alpha}
\alpha=\frac{U_{12}}{\sqrt{U_{11}U_{22}}}
\end{equation}
is crucial in qualitatively determining the TF ground state. If
$\alpha>1$, which is the case of interest here, the ground state
density profiles are completely separated in adjacent regions
divided by domain walls. In the regions where only species $k$ is
present, the solution to the TF equations reads
\begin{equation}\label{tf}
\rho_k^{\TF}(x)=\frac{\mu_k-V_k(x)}{U_{kk}}.
\end{equation}
Equation~(\ref{tf}) completely determines  the functional
dependence of the densities on the external potential (and on the
chemical potentials), once their supports are given. Here we are
interested in the ground state solution, so that the supports are
determined by minimizing the energy of the system~(\ref{engcpsi})
as a function of the number and the positions of the domain walls.
In Ref.~\cite{binbec}  we proved that for  continuous trapping
potentials stationarity of energy requires that the densities at a
domain wall at $R_j$ satisfy
\begin{equation}\label{rhowall}
\sqrt{U_{11}}\rho_1(R_j)=\sqrt{U_{22}}\rho_2(R_j).
\end{equation}
Moreover, in the special case $V(x)\equiv V_1(x)=V_2(x)$,
conditions (\ref{rhowall}) imply that the external potential, and
thus the density of each species, should be the same at all domain
walls.

The TF approximation works very well for large numbers of
particles, and provides a good estimate of the energy of the
system. Despite being small with respect to the potential energy,
corrections due to the kinetic term give nonetheless rise to
macroscopic effects. The most relevant of such effects is the
crossing between stationary states with different numbers of
domain walls in the ground state \cite{kasamatsu,trippenbach}. It
is thus necessary, in order to determine the actual ground state
of a mixture, to consider a regularization scheme of the TF
density profiles, that enables to smooth parts of the TF profiles,
like domain walls and zeros, that provide an infinite contribution
to the kinetic energy \cite{stringari}.

\section{Variational regularization of density profiles}
\label{sec:reg}

Many attempts have been made so far in order to consistently
estimate the energy corrections due to the kinetic contribution.
The pioneering works by Ao and Chui \cite{aochui} and Timmermans
\cite{timm} are based on an exponential approximation of the  TF
density profiles, by extrapolating the solution of the
Gross-Pitaevskij  equations far from a domain wall, and on a
linear approximation of the regularized walls, respectively. Other
authors \cite{barankov,van} put forward a regularization with
fixed chemical potential in various regimes.

The task we will try to accomplish in this work is to find a
proper domain wall regularization, which provides a reliable
approximation to the ground state profile and energy of a binary
mixture in a trapping potential, and which is at the same time an
analytically manageable trial function. The approximations in
\cite{aochui,timm} are based on rather crude ansatz, but provide a
good estimation of the order of magnitude of the energy changes
due to the kinetic terms. However, we will try to find better
upper bounds to the ground-state energies, since it is possible
that a small energy change results in macroscopic differences in
density profiles. Our approximation will be referred to a system
with fixed numbers of particles, and will strictly rely on the
preservation of the normalization conditions. It is indeed
difficult to use results obtained with fixed chemical potential
\cite{barankov,van} in this case, since for non trivial external
potentials it is generally impossible to invert the normalization
conditions (\ref{normal}) and explicitly express the chemical
potentials as functions of the numbers of particles.

\subsection{Domain walls}

In order to regularize the domain walls, it is sufficient to
replace the singular TF density profiles with a continuous
function, with bounded first derivative. The minimization of the
kinetic energy leads to tails of each species penetrating through
the domain wall.

\begin{figure}
\centering
\includegraphics[width=0.45\textwidth]{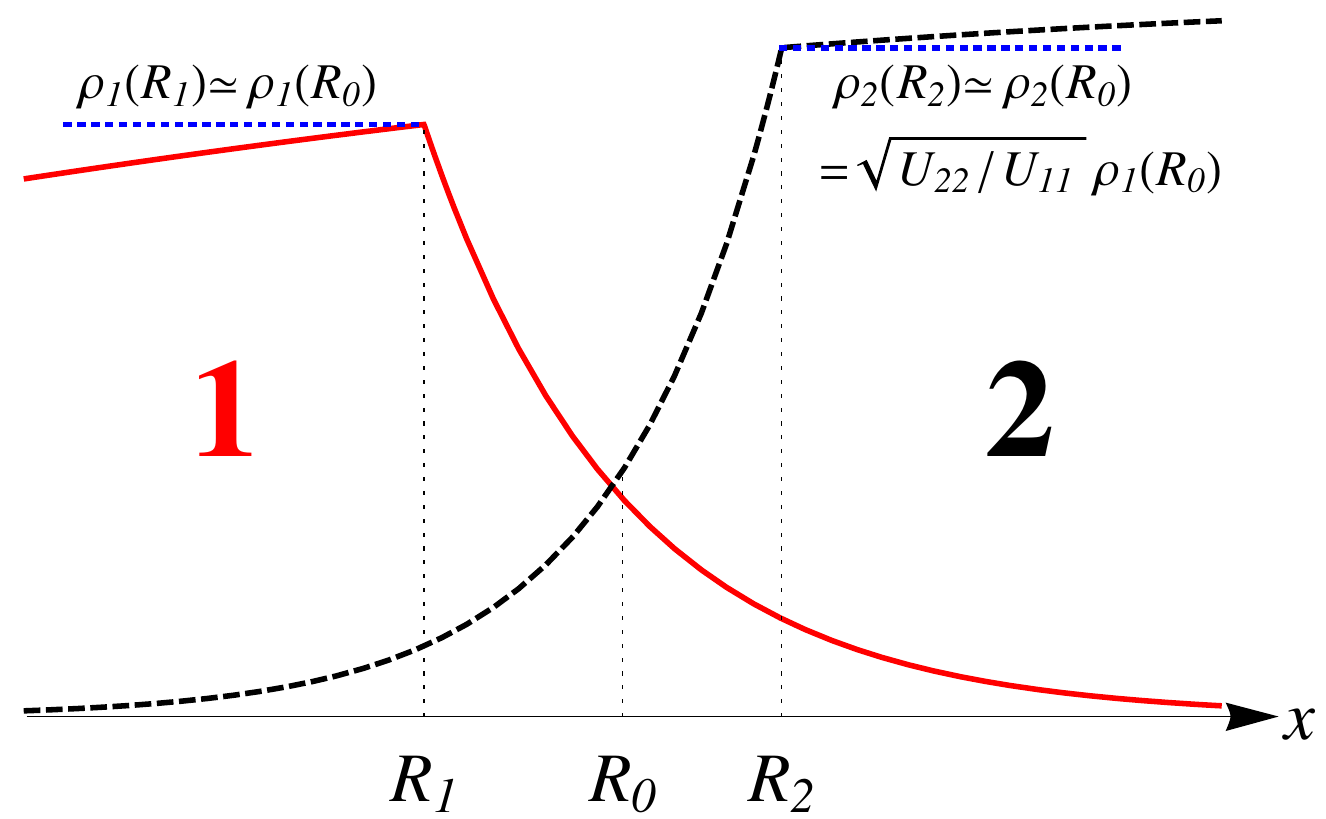}
\caption{(Color online) Plot of the trial density profiles
(\ref{tilderho1})-(\ref{tilderho2}) in a neighborhood of a domain
wall placed at $R_0$. The solid (red) line represents the density
of the first species, the dashed (black) line the density of the
second species, with $U_{11}<U_{22}$.}\label{fig:regular}
\end{figure}

Our choice of trial profiles is based on an exponential tail
regularization of the TF solutions. Let us consider a domain wall
placed at position $R_0$. In its neighborhood we assume as
profiles the continuous functions
\begin{equation}\label{tilderho1}
\tilde{\rho}_1(x)= \left\{\begin{array}{lcr}
\rho_1^{\mathrm{TF}}(x) & \quad & \text{if}\quad x<R_1 \\
\rho_1^{\mathrm{TF}}(R_1)\e^{-2(x-R_1)/\Lambda_1} & \quad &
\text{if}\quad x\geq R_1
\end{array}\right.
\end{equation}
and
\begin{equation}\label{tilderho2}
\tilde{\rho}_2(x)= \left\{\begin{array}{lcr}
\rho_2^{\mathrm{TF}}(x) & \quad & \text{if}\quad x>R_2 \\
\rho_2^{\mathrm{TF}}(R_2)\e^{2(x-R_2)/\Lambda_2} & \quad &
\text{if}\quad x\leq R_2
\end{array}\right. ,
\end{equation}
with $R_1 < R_0 <R_2$, $\Lambda_k>0$, and $\rho_k^{\mathrm{TF}}$
the TF densities~(\ref{tf}). See Fig.~\ref{fig:regular}.
The points $R_k$ are fixed in order to locally preserve the
normalization conditions. This is accomplished by imposing that
the integral of the removed part of the TF profiles be equal to
that of the new exponential tails:
\begin{equation}
\int_{R_k}^{R_0} dx\, \rho_k^{\TF}(x)=
\int_{R_k}^{(-1)^{k+1}\infty} dx\, \tilde{\rho}_k(x),
\end{equation}
which gives
\begin{equation}\label{Rk}
R_k-R_0=(-1)^k \frac{\Lambda_k}{2}\left[1+\Or\left( \Lambda_k
\frac{\rho_k^{\TF\prime}(R_0)}{\rho_k^{\TF}(R_0)} \right) \right].
\end{equation}
In the following, we will neglect all corrections depending on the
first derivatives of the TF densities, $\rho_k^{\TF\prime}= d
\rho_k^{\TF}/ dx$, which depend  linearly on the derivatives of
the external potentials. This assumption, which involves
conditions on the (by now arbitrary) parameters $\Lambda_k$, will
be expressed at the end of our calculations in term of physical
quantities. It is remarkable that under this approximation the
regularized density profiles still satisfy condition
(\ref{rhowall}) at $R_0$.

Once the trial profiles are chosen, we proceed to the computation
of the difference in potential and interaction energy with respect
to the TF densities. It is convenient to explicitly show the
dependence of the energy on the largest penetration length, say
$\Lambda_1$, and the ratio $\eta=\Lambda_2/\Lambda_1\leq 1$. Since
$\alpha>1$, the mixing must result in an increase in potential
energy. The self interaction energy associated to the exponential
tail of $\tilde{\rho}_1(x)$ reads
\begin{equation}
\tilde{\U}_{11}=\frac{U_{11}}{2} \int_{R_1}^{\infty}
\tilde{\rho}_1(x) \, dx= \frac{1}{8}
\Lambda_1(\rho_1^{\TF}(R_0))^2.
\end{equation}
This contribution replaces the self interaction energy of the
removed TF density:
\begin{equation}
\U_{11}^{\TF}=\frac{U_{11}}{2} \int_{R_1}^{R_0} \rho_1^{\TF}(x) \,
dx= \frac{1}{4} \Lambda_1(\rho_1^{\TF}(R_0))^2.
\end{equation}
The same results hold for the second species after the
substitution $\Lambda_1\to\eta\Lambda_1$. As expected, extending
the density profiles implies a reduction in the self interaction
energy, $\U_{\mathrm{self}} = \U_{11} + \U_{22}$, which reads
\begin{equation}\label{Uself}
\Delta\U_{\mathrm{self}}(\Lambda_1,\eta)= -\frac{1}{8}(1+\eta)
U_{11} \Lambda_1 (\rho_1^{\TF}(R_0))^2 .
\end{equation}
A positive contribution comes from the inter-species interaction
terms, $\U_{\mathrm{inter}} = \U_{12} + \U_{21}$, which are due to
the penetration of the tails in the bulk of the other species and
their superposition around $R_0$. (Remember that in a  TF
separated configuration $\U_{\mathrm{inter}} = 0$.) The total
change in the inter-species interaction energy reads
\begin{eqnarray}
\label{Uinter} \U_{\mathrm{inter}}(\Lambda_1,\eta) &=& U_{12}
\int_{-\infty}^{\infty} \tilde{\rho}_1(x) \tilde{\rho}_2(x) \, dx
\nonumber
\\ &=& \frac{ \alpha U_{11} \Lambda_1 \left( \mathrm{e}^{-(
1+\eta)} - \eta^2 \mathrm{e}^{-(
1+1/\eta)} \right) }{2(1-\eta)} \nonumber \\
& & \times (\rho_1^{\TF}(R_0))^2.
\end{eqnarray}
Observe that the limit $\eta\to1$ is finite. It is easy to verify
that no corrections come from the interaction with the external
potential at the chosen order of approximation. The results
(\ref{Uself})-(\ref{Uinter}) are found under the hypothesis that
the distance separating the considered domain wall from other
possible walls is much larger than the $\Lambda_k$'s.

Let us now consider the contributions from the kinetic energy. The
value of the kinetic energy of the densities in the bulk is
consistently neglected in our approximation, being $\Or(
(\Lambda_k\rho_k^{\TF\prime}/\rho_k^{\TF})^2)$ with respect to the
leading terms. On the other hand, the contribution across the
domain wall depends on the inverse penetration lengths and reads
\begin{eqnarray}\label{Kinetic}
\K_{\mathrm{wall}}(\Lambda_1,\eta) & = & \frac{\hbar^2}{2m_1}
\int_{R_1}^{+\infty} \left|\frac{d\sqrt{\tilde{\rho}_1}}{dx}
\right|^2 dx \nonumber \\ & + &   \frac{\hbar^2}{2m_2}
\int_{-\infty}^{R_2} \left|\frac{d\sqrt{\tilde{\rho}_2}}{dx}
\right|^2 dx \nonumber \\
& = & \frac{\hbar^2}{4 m_1 \Lambda_1}  \left(
1+\frac{\eta_0^2}{\eta} \right) \rho_1^{\TF}(R_0),
\end{eqnarray}
where
\begin{equation}
\eta_0 \equiv \frac{\xi_2}{\xi_1} = \left( \frac{m_1}{m_2}
\right)^{\frac{1}{2}} \left( \frac{U_{11}}{U_{22}}
\right)^{\frac{1}{4}}
\end{equation}
is the ratio between the healing lengths
\begin{equation}
\xi_k=\hbar/\sqrt{2 m_k U_{kk} \rho_k^{\TF}(R_0)},
\end{equation}
$(k=1,2)$ of uniform condensates whose densities are
$\rho_k^{\TF}(R_0)$ \cite{stringari}.

The energetic contributions (\ref{Uself})--(\ref{Kinetic}) depend
on the free parameters $\Lambda_k$. Heuristically, as a benchmark,
one can evaluate them at $\Lambda_k = \xi_k/\sqrt{\alpha-1}$, the
penetration lengths associated with the exponential tail of one
species into the bulk of the other one, which are obtained via an
approximated Gross-Pitaevskij  equation \cite{aochui}. Besides
yielding a divergence of the penetration for $\alpha$ close to 1,
which is unphysical for a system with finite numbers of particles,
this is not the optimal choice, and it can even yield an energy
20\% larger than the minimum for physical parameters. In a
situation where small energy changes are involved, this
discrepancy can be very relevant. The best strategy is  to
minimize the total energy of the wall over the family of trial
functions~(\ref{tilderho1})-(\ref{tilderho2}) parametrized by
$\Lambda_1$ and $\eta=\Lambda_2/\Lambda_1$:
\begin{equation}\label{Etot}
\mathcal{E}_\textrm{wall} (R_0) = \min_{\Lambda_1,\eta}
\left\{\Delta\U_{\mathrm{self}}+\U_{\mathrm{inter}}+\K_{\mathrm{wall}}\right\}.
\end{equation}
This will enable us to get a much more accurate upper bound for
the ground state energy of the binary mixture.

\begin{figure}
\centering
\includegraphics[width=0.45\textwidth]{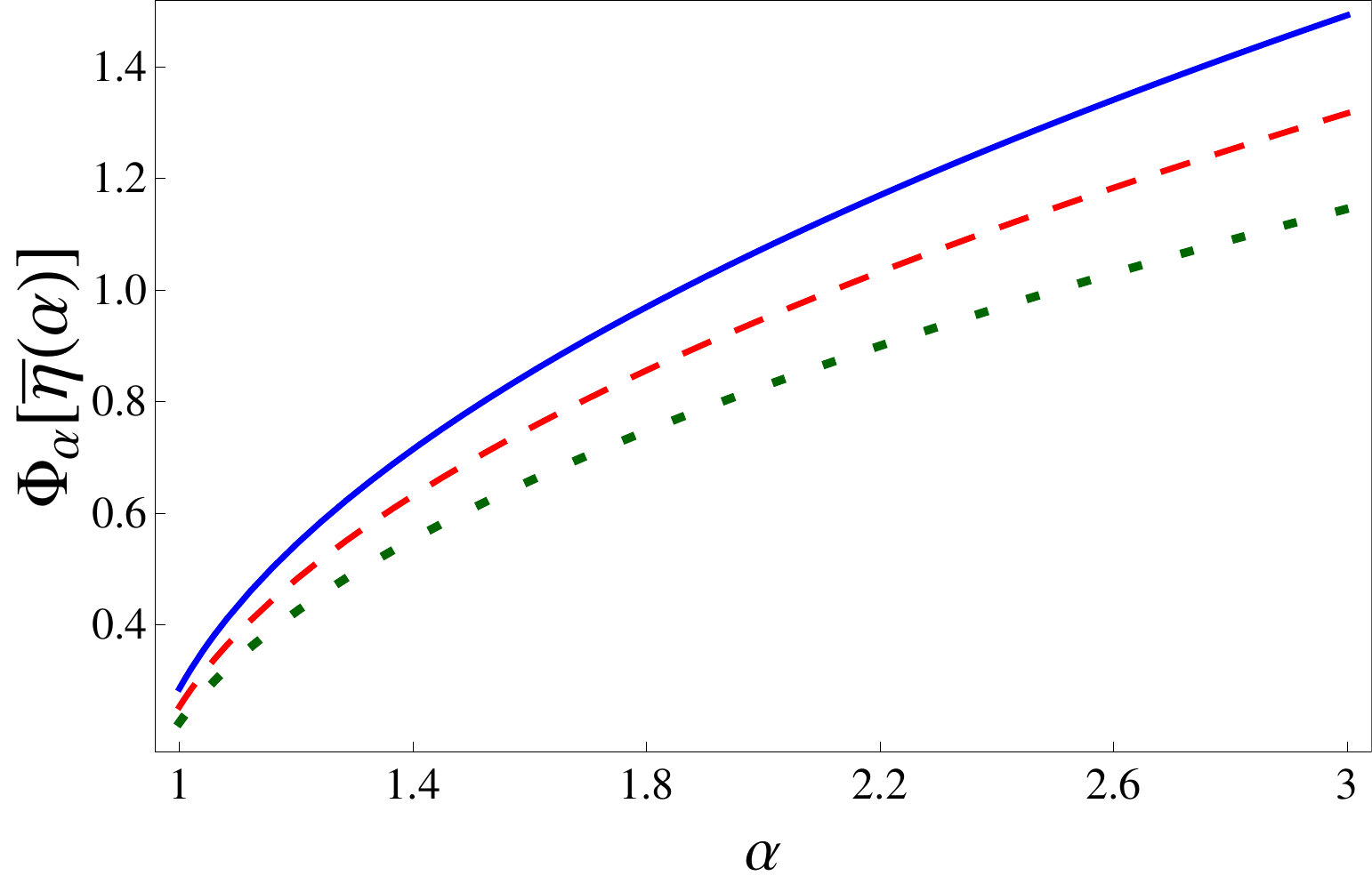}
\caption{(Color online) Behavior of the function
$\Phi_{\alpha}(\bar{\eta}(\alpha) )$, obtained by solving Eq.\
(\ref{minim}), for different values of $\eta_0$. Solid (blue)
line: $\eta_0=0.99$; dashed (red) line: $\eta_0=0.75$; dotted
(green) line: $\eta_0=0.5$.} \label{fig:Phi}
\end{figure}

\begin{figure}
\centering
\includegraphics[width=0.45\textwidth]{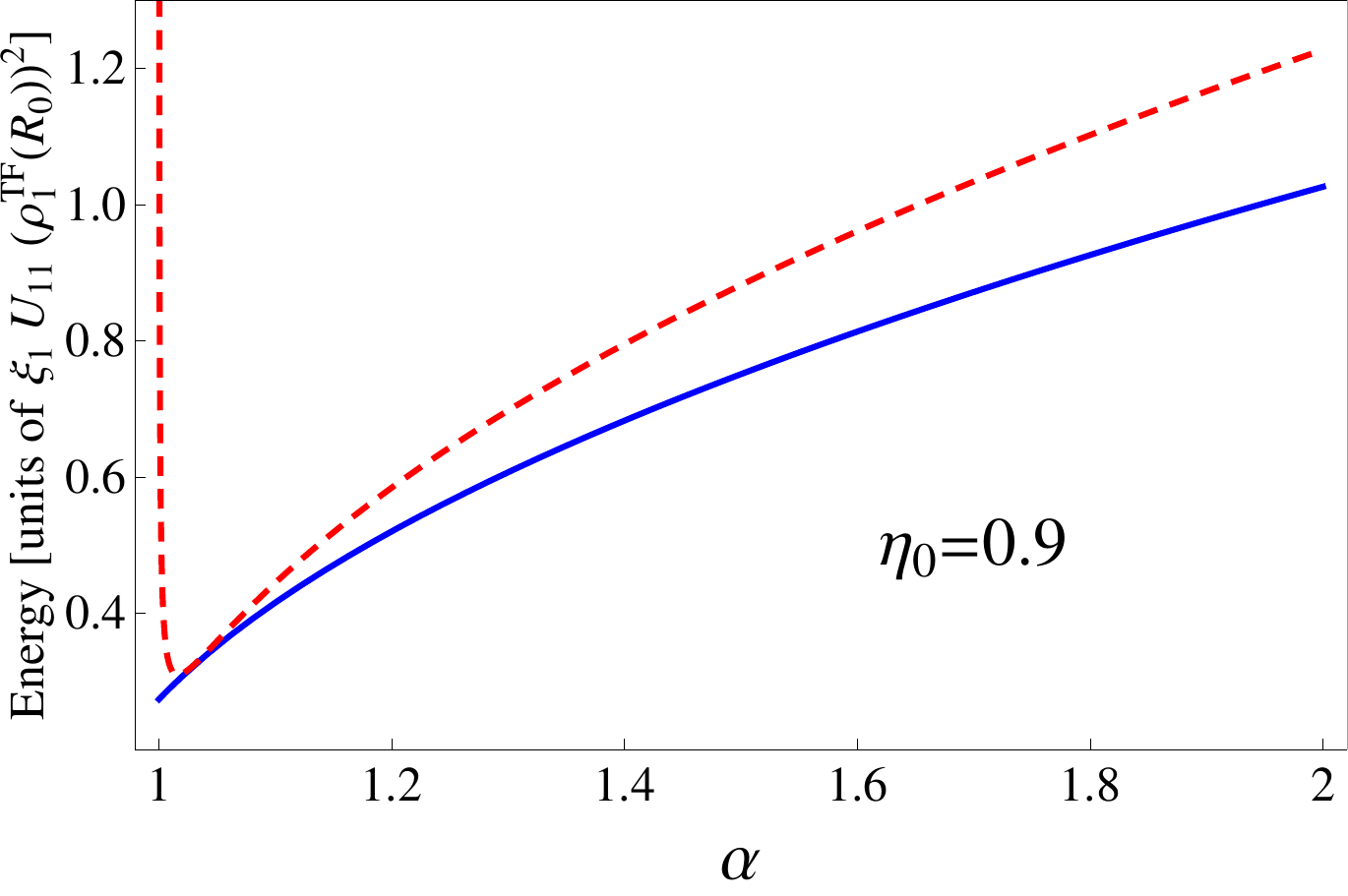}
\caption{(Color online) Comparison between the minimum energy
[units of $\xi_1 U_{11} (\rho_1^{\TF}(R_0))^2$] of a domain
wall~(\ref{Ewall}) (solid blue line) and the energy computed for
penetration lengths $\xi_k/\sqrt{\alpha-1}$, for $\eta_0=0.9$
(dashed red line).} \label{fig:comp}
\end{figure}

\begin{figure}
\centering
\includegraphics[width=0.45\textwidth]{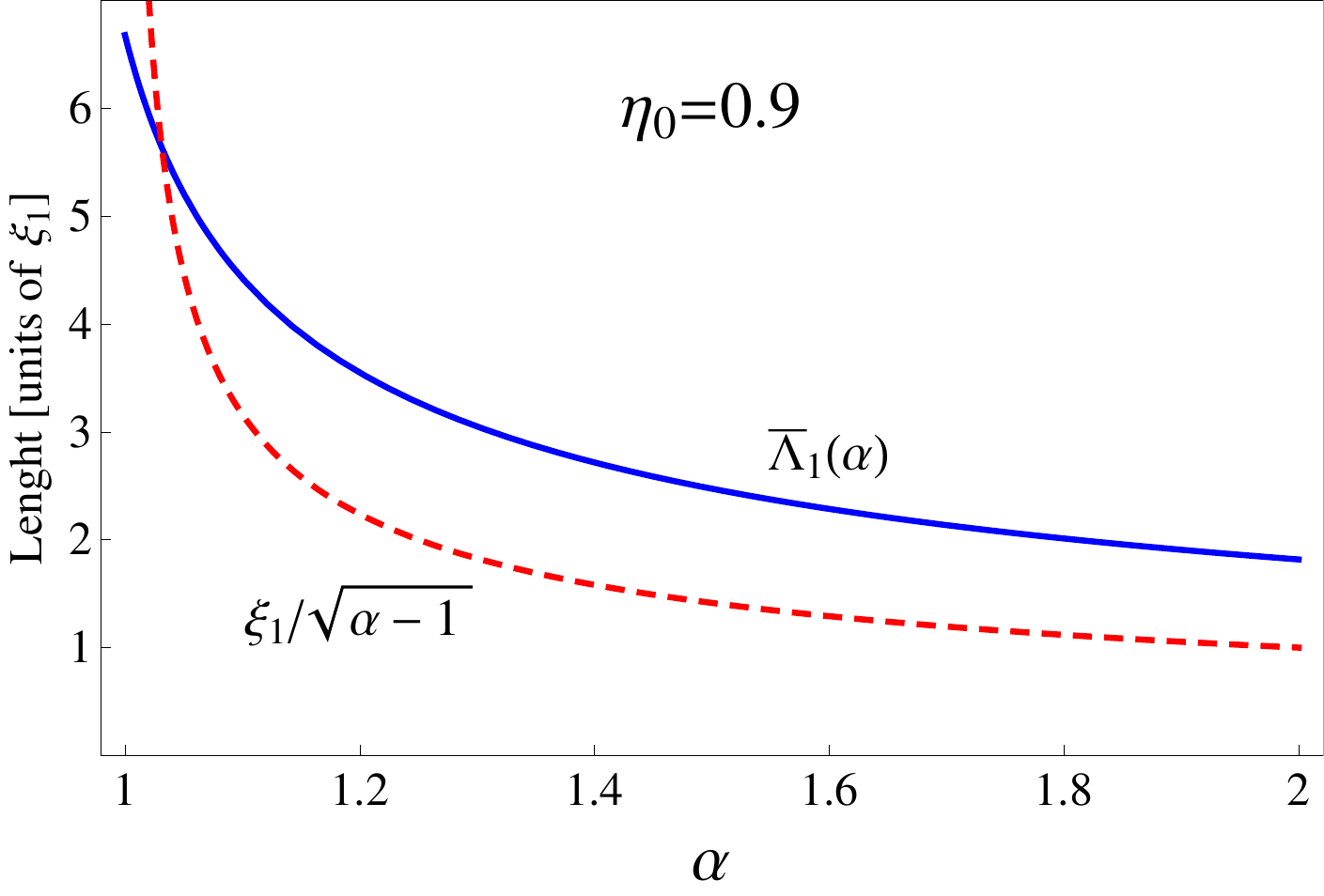}
\caption{(Color online) The solid (blue) line represents the
optimal penetration length $\bar{\Lambda}_1(\alpha)$ (units of
$\xi_1$) plotted against $\alpha$ for $\eta_0=0.9$. The dashed
(red) line is the bulk penetration length $\xi_1/\sqrt{\alpha-1}$
\cite{aochui}, plotted here for comparison.}\label{fig:Lambda}
\end{figure}

The minimum in~(\ref{Etot}) is attained at a single point
$(\bar{\Lambda}_1(\alpha),\bar{\eta}(\alpha))$ and reads
\begin{eqnarray}\label{Ewall}
\mathcal{E}_{\mathrm{wall}}(R_0)= \left(\frac{\hbar^2 U_{11}
(\rho_1^{\TF}(R_0))^3 }{2 m_1}\right)^{\frac{1}{2}} \Phi_{\alpha}(
\bar{\eta}(\alpha) ),
\end{eqnarray}
where
\begin{eqnarray}\label{Phi}
\Phi_{\alpha}(\eta) &=& \left( \alpha \frac{(\e^{-(1+\eta)} -
\eta^2 \e^{-(1+1/\eta)})}{1-\eta}-\frac{1+\eta}{4} \right)^{1/2}
\nonumber\\
& &\times  \left(1+\frac{\eta_0^2}{\eta} \right)^{1/2}.
\end{eqnarray}
 The optimal ratio between the penetration
lengths, $\bar{\eta}(\alpha)$, is the solution to the
transcendental equation
\begin{eqnarray}
& & \e^{\eta} [\e^{1 + 1/\eta} (1 - \eta)^2 (\eta^2 - \eta_0^2) +
4 \alpha \, \eta (\eta + \eta^2 - \eta^3 + \eta_0^2) ] \nonumber
\\ & & \qquad =
4 \alpha \, \e^{1/\eta} (\eta^3 - (1 - \eta - \eta^2) \eta_0^2)  .
\label{minim}
\end{eqnarray}
In Fig.\ \ref{fig:Phi} the function $\Phi_{\alpha}(
\bar{\eta}(\alpha) )$ is plotted versus $\alpha$ for different
values of the parameter $\eta_0$, while in Fig.\ \ref{fig:comp} a
comparison is displayed between the minimized energy and the
result computed with the bulk penetration lengths
$\xi_k/\sqrt{\alpha-1}$. It can be observed that the minimizing
energy has a monotonic behavior with $\alpha$.
The optimal penetration length
\begin{equation}\label{Lambda}
\bar{\Lambda}_1(\alpha)=\xi_1 \frac{\left( 1+
\frac{\eta_0^2}{\bar{\eta}(\alpha)} \right)}{\Phi_{\alpha}\left(
\bar{\eta}(\alpha) \right)},
\end{equation}
is plotted versus  $\alpha$ (for $\eta_0=0.9$) in Fig.\
\ref{fig:Lambda}.
The conditions ensuring that terms depending on the first
derivatives of the densities can be neglected are thus summarized
by the following inequality
\begin{equation}\label{appr}
\frac{1+\eta_0^2/\bar{\eta}(\alpha)}{\Phi_{\alpha}(\bar{\eta}(\alpha))}
\frac{\hbar}{\sqrt{2m_1}}
\frac{|V'(R_0)|}{U_{11}(\bar{\rho}_1)^{3/2}} \max_{j,k}
\frac{\sqrt{U_{jj}}}{U_{kk}} \ll 1.
\end{equation}
This condition also ensures that the distance between domain walls
is always larger than $\Lambda_k$'s, thus validating the results
in Eqs.\ (\ref{Uself})-(\ref{Uinter}).

Since the dependence on $\alpha$ and $\rho_1^{\TF}$ in
(\ref{Ewall}) is factorized, the total correction for a TF
configuration with domain walls placed at $\{R_1,\dots,R_w\}$ is
\begin{equation}\label{walltotal}
\Delta\mathcal{E}^{(w)}=\sum_{j=1}^w
\mathcal{E}_{\mathrm{wall}}(R_j)\equiv C_{\alpha} \sum_{j=1}^w
(\rho_1^{\TF}(R_j))^{3/2},
\end{equation}
with
\begin{equation}
C_{\alpha}=\hbar(U_{11}/2m_1)^{1/2}\Phi_{\alpha}(\bar{\eta}(\alpha)).
\end{equation}

For  confining potentials proportional to each other, including
the interesting case in which they are equal, the energy increase
(\ref{Ewall}) is independent of the specific domain wall, since
(\ref{rhowall}) implies that the densities are the same at all the
walls of a stationary configuration, namely $\rho_1^{\TF}(R_j)=
\rho_1^{\TF}(R_0)$ for any $j$ \cite{binbec}. This implies that
the energy correction to a TF configuration with $w$ domain walls
is simply
\begin{equation}
\Delta\mathcal{E}^{(w)}= w \mathcal{E}_{\mathrm{wall}}(R_0).
\end{equation}

On the other hand, the condition of applicability of our
approximations (\ref{appr}) depends on the specific domain wall
through the first derivative of the potentials at each wall. Of
course, a uniform control on the derivatives of the potentials
would give a sufficient condition for their applicability to all
possible configurations.

It shoud be emphasized that the validity of our approximation has
an upper bound in $\alpha$. Indeed, it can be deduced from Eq.\
(\ref{Phi}) and Fig.\ \ref{fig:comp} that the function
$\Phi_{\alpha}(\bar{\eta}(\alpha))$ is not bounded from above as
$\alpha$ increases, leading to a divergent correction to the TF
energy, which frustrates regularization attempts. The reason lies
in the fact that for $\alpha\to\infty$ the exponential tail ansatz
is no longer justified, since in this case one species feels the
other one like a hard wall, thus leading to solutions of the form
$\tanh(x/\sqrt{2}\xi_k)$ and to a saturation of the domain wall
energy. Since the typical variation lengths of the densities are
in this case $\sqrt{2}\xi_k$, we can bind the validity of our
ansatz to values of $\alpha$ verifying
$\bar{\Lambda}_k(\alpha)\lesssim \sqrt{2}\xi_k$ (see Fig.\
\ref{fig:Lambda} for a comparison), typically corresponding to
values $\alpha\lesssim 2.5\div 3$, which matches well the
experimental ranges.

\subsection{Profile edges}

The presence of first order zeros at the edges of the TF density
profiles, which are compactly supported, gives rise to logarithmic
divergencies in the kinetic energy. Even this situation can be
tackled by a proper regularization of the densities, leading to an
increase in the potential energy and to a  kinetic contribution
\cite{stringari}. To this end, we consider a zero at position
$x_0$ and conventionally consider a TF profile
$\rho_k^{\TF}(x)\theta(x_0-x)$ in a neighborhood of $x_0$,
$\theta$ being the unit step function. The regularization is based
on the solution of the linearized single-condensate wave function
in a neighborhood of $x_0$, namely $\sqrt{\rho_k} \propto
\Ai(x/\delta)$, with $\Ai(y)$ the proper Airy function
\cite{abramowitz}, decreasing as $y\to\infty$, and
\begin{equation}
\delta_k = \left( \frac{2m_k F_k(x_0)}{\hbar^2} \right)^{-1/3},
\quad \text{with } F_k(x_0)=\left| \frac{dV_k}{dx} \right|_{x_0},
\end{equation}
the characteristic length. Corrections to a solution based on the
linear approximation of the potential are negligible if the second
derivative of the potential is much smaller than $(2m_k
F_k^4(x_0)/\hbar^2)^{1/3}$.

Following such scheme, we consider the trial family of (continuous
and positive) regularized TF profiles
\begin{equation}\label{regzero}
\tilde{\rho}_k(x)= \left\{\begin{array}{lcr}
\rho_k^{\mathrm{TF}}(x) & \quad & \text{if}\quad x< x_k \\
\rho_k^{\mathrm{TF}}(x_k)\,  f\left( \frac{x-x_k}{\delta_k}
\right)^2 & \quad & \text{if}\quad x\geq x_k
\end{array}\right. ,
\end{equation}
with $f(y)= \Ai(y)/ \Ai(0)$. As in the case of the domain wall,
the point $x_k$ is determined by requiring the local normalization
condition to be fulfilled. The result, analogous to (\ref{Rk}),
reads
\begin{equation}\label{Pk}
x_k-x_0= - 2 I_0 \delta_k,
\end{equation}
where $I_0=\int_{0}^{\infty}dy f^2(y)\simeq 0.53 $. Summing the
variations of the self-interaction energy (negative) and of the
external potential energy (positive) to the  kinetic energy yields
the total energy change due to the regularization of the zero of
the density profile
\begin{equation}\label{Ezero}
\mathcal{E}_{\mathrm{zero}}^{(k)}(x_0) \simeq 0.274\,
\frac{\hbar^2}{m_k U_{kk}} F_k(x_0).
\end{equation}
This contribution depends on the specific zero, as well as on the
species $k$, since the first derivatives of the potential at its
zeros are generally not related.

In order to get a feeling for the orders of magnitude of the
various energies, one can consider the simple case in which the
trapping potential is well approximated by a power law
\begin{equation}
V(x)\sim |x|^n.
\end{equation}
A TF zero, placed at $x_0$, of the density profile of species $k$,
is determined by the condition $\mu_k=V(x_0)\sim x_0^n$. By
considering the last relation and the normalization conditions,
one can obtain the scaling law of the chemical potential with
respect to the number of particles, namely $\mu_k\sim
N_k^{n/(n+1)}$. Hence one can obtain the scaling laws of the TF
energy, $\mathcal{E}_{\TF} \sim N_k^{(2n+1)/(n+1)}$, and of the
kinetic energy, including the contribution  (\ref{Ezero}),
$\mathcal{T} \sim N_k^{(n-1)/(n+1)}$. The energy of a domain wall
depends on $(\rho^{\TF})^{3/2}\sim \mu^{3/2}$, and thus
$\mathcal{E}_{\mathrm{wall}}\sim N_k^{3n/(2n+2)}$. The higher
order corrections  of $\Or(\bar{\Lambda}_k \rho_k'/\rho_k)$ terms
accidentally scale like the kinetic energy. Thus, in a
configuration with $w$ domain walls, the trial ground state has
energy
\begin{equation}
\mathcal{E}=\mathcal{U}_{\TF}+ w  \mathcal{E}_{\mathrm{wall}} +
\Or\left(N_k^{\frac{n-1}{n+1}}\right),
\end{equation}
where
\begin{equation}
\mathcal{U}_{\TF}= \mathcal{E}_{\TF}(\rho_1^{\TF},\rho_2^{\TF})
\end{equation}
is the energy of the TF densities.

\section{Domain wall suppression: a case study}
\label{sec:dw}

As an application of the previous results we consider in this
section the energy crossing between configurations with different
numbers of domain walls in a double well potential. We will
consider a physical situation in which a crossing between ground
states with a maximal and a minimal number of domain walls can be
observed. We introduce an operational way to control the crossing,
based on a scaling property of the TF energy functional, thus
suggesting a possible experimental realization.

Notice first that the effect of the kinetic energy on the ground
state of a mixture in a square well potential is trivial, since in
this case the TF energy for separated configurations depends only
on the volumes occupied by the two species and not on how they are
distributed inside the well. Therefore, inclusion of the domain
wall energy immediately enables us to identify the configuration
with a single domain wall as the ground state.

The situation is much more interesting for potentials that vary
over the region occupied by the mixture. In the TF theory, density
profiles with a maximal number of domain walls are usually
energetically favored, especially when the ratio of the
self-interaction coefficients is very close to one \cite{binbec}.
However, the inclusion of the kinetic energy can drastically
change this picture. Each domain wall has an energetic cost,
expressed by (\ref{Ewall}), whose effect on the total energy
decreases as the numbers of particles increase. Thus, for very
large number of particles, the ground state is more likely to have
a maximal number of domain walls, but if the number of particles
decreases or the $\alpha$ parameter increases, it can become more
convenient to reduce the number of walls.

As a test ground for the effectiveness of our method, let us
consider an example of this phenomenon, that was analyzed by
numerical integration of the coupled Gross-Pitaevskii stationary
equations in Ref.\ \cite{kasamatsu}. The potentials are harmonic
and the possible competing ground states have one or two walls: a
mixture of $^{87}$Rb atoms with $m_1=m_2=m=1.45\times 10^{-25}$ kg
in two hyperfine states was considered, with scattering lengths
$a_1=5.36$ nm and $a_2=5.66$ nm. The longitudinal trapping
frequency was fixed to $\omega=2\pi\times 90$ Hz, with the
transverse trapping frequency 30 times larger. The authors were
able to build a phase diagram showing the crossing between single-
and double-wall configurations by applying numerical techniques.
An explicit comparison of the results for the total energies of
the configurations are given in \cite{kasamatsu} for
$N_1=N_2=2000$ and $\alpha=1.18$.

In Table \ref{table} we compare our results with those of Ref.\
\cite{kasamatsu}. It is manifest that, while for the
symmetry-preserving (double-wall) state the two results are
identical up to the fourth significant digit, our regularization
method enables us to attain a stricter upper bound for the
ground-state energy of the symmetry-breaking (single-wall)
configuration. Since the choice of  the trial densities in the
energy functional is based on the physics of the phenomenon, it is
not surprising that our analytical regularization technique,
together with the exact results coming from TF, leads to a better
approximation of the ground state of a binary mixture than the
accurate numerical integration of the coupled Gross-Pitaevskii
equations \cite{kasamatsu}.

\begin{table*}
\caption{Total energy of a binary mixture of $^{87}$Rb atoms in a
harmonic potential with longitudinal frequency $\omega=2\pi\times
90$ Hz and transverse frequency 30 times larger, for
$N_1=N_2=2000$ and $\alpha=1.18$. The results obtained with the
analytical approximation schemes proposed in this article are
compared with those numerically obtained in Ref.\
\cite{kasamatsu}.} \label{table}
\begin{tabular}{ | c |c | c|}
\hline
& One wall (symmetry breaking) &  Two walls (symmetry preserving) \\
\hline
$\U_{\TF}$ & $87.091$   & 86.772 \\
$\mathcal{E}= \U_{\TF} +  \Delta \mathcal{E}^{(w)}$  &  87.486   &  87.423  \\
$\mathcal{E}$ numerically computed in \cite{kasamatsu}  &  87.551   &  87.426 \\
\hline
\end{tabular}
\end{table*}

\begin{figure}
\centering
\includegraphics[width=0.45\textwidth]{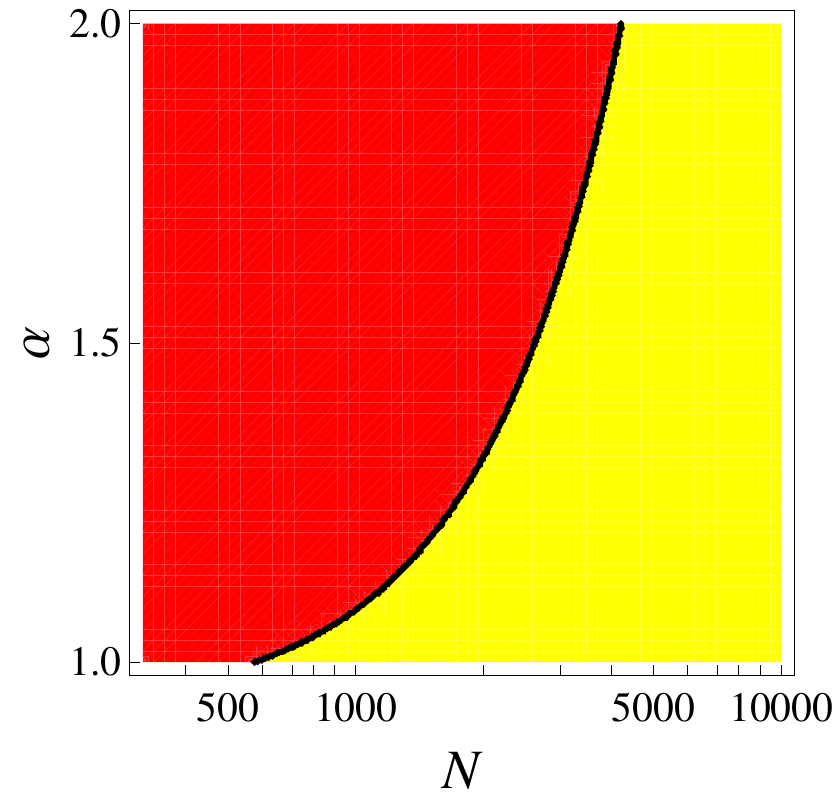}
\caption{(Color online) Ground state phase diagram in the
$N$--$\alpha$ plane for a binary mixture of $^{87}$Rb atoms in a
harmonic potential with longitudinal frequency $\omega=90$Hz and
transverse frequency 30 times larger. In the light grey region
(yellow in the online version) the ground state is a symmetry
preserving configuration with two domain walls, while in the dark
grey region (red in the online version) a symmetry breaking
configuration, with a single domain wall, is energetically
favored.}\label{fig:harmonic1}
\end{figure}

The approximation on the energy of the trial densities was proved
to be robust by a numerical check, in which the  total energy of
the regularized TF profiles is computed by numerical integration,
showing only a slight increase in the fifth significant digit. The
differences in the estimate of the ground-state energy of the
symmetry breaking configurations, which are general and not
restricted to the aforementioned case, lead to a different phase
diagram in the plane $(N_1=N_2=N,\alpha)$, shown in Fig.\
\ref{fig:harmonic1}, that should be compared with that in Fig.\ 2
of Ref.\ \cite{kasamatsu}. In our case the transition line is
shifted by a factor $\simeq 1.5$ with respect to the $N$ axis
(towards larger values of $N$). Thus, according to our analysis,
the symmetry breaking ground state is present in a larger region
of the $(N,\alpha)$ plane, where it was previously not expected.

\subsection{Scaling properties}

It is convenient to study the scaling properties of the energy
terms in (\ref{engcpsi}) under a dilation. If lengths scale as
$x\to x/a$ with $a>0$, it is easy to see from~
(\ref{normal})-(\ref{engcpsi1}) that the kinetic energies scale as
$\mathcal{T}_k \to \mathcal{T}_k/a$, while the potential and
interaction energies scale as $\mathcal{V}_k\to a \mathcal{V}_k$
and $\U \to a \U$, and accordingly the numbers of particles scale
as $N_k \to a N_k$. Therefore, the larger $a$ and the numbers of
particles, the smaller the ratio between kinetic and potential
energy. Let us look at this property in more details.

Consider TF density profiles $\rho_k^{\TF}$ in a
separated configuration given by~(\ref{tf}) 
and fix their supports  by choosing the number $w$ of domain walls
and their positions $R_j$, all satisfying $V(R_j)=v$ with $v$ a
constant value. Let the density profiles be normalized to $N_k$.
The TF potential energy of the separated configuration reads
\begin{eqnarray}
& &\mathcal{U}_{\TF}(N_k,V,w)
\nonumber\\
& & = \sum_{k=1,2} \int_{\Omega_k} dx\,
\left(V(x)\rho_k^{\TF}(x)+\frac{U_{kk}}{2}\rho_k^{\TF}(x)^2
\right). \quad
\end{eqnarray}
The integration domains $\Omega_k=\Omega_k(\{R_j\},\{P_j^{(k)}\})$
are unions of intervals whose ends are  domain walls or edges,
located at $\{R_j\}=V^{-1}(\{v\})$ and
$\{P_j^{(k)}\}=V^{-1}(\{\mu_k\})$, respectively. Observe now that
if the potential is scaled as
\begin{equation}
\label{scaleW} W(x)=V(x/a),
\end{equation}
while leaving unchanged the chemical potentials $\mu_k$ and the
domain wall potential $v$, the density profiles
$\sigma_k^{\TF}(x)=\rho_k^{\TF}(x/a)$ are still TF solutions,
corresponding to the potential $W$, to supports $a \Omega_k
=\Omega_k(\{a R_j\},\{a P_j^{(k)}\})$, and to numbers of particles
\begin{equation}
\int_{a\Omega_k} dy\,\sigma_k^{\TF}(y)= a \int_{\Omega_k} dx
\rho_k^{\TF}\left(x\right)=aN_k.
\end{equation}
Moreover, the energy of the scaled configuration is related to the
previous one by
\begin{equation}
\U_{\TF}(aN_k,W,w)=a \, \U_{\TF}(N_k,V,w).
\end{equation}
On the other hand, the energy contribution of the domain walls is
unchanged by the scaling, since it depends only on fixed
quantities, namely the number of walls, the chemical potentials
and the domain wall potential.

\subsection{Energy crossing}

Let us consider for definiteness a physical system with equal
numbers $N_0$ of particles of the two species in a potential $V$,
with $w$ domain walls, potential energy $\U_{\TF}(N_0,V,w)$ and
total domain wall correction $\Delta\mathcal{E}^{(w)}$,
proportional to $w$. If the numbers of particles are increased to
$N = a N_0$ with $a>1$ and the potential is stretched to
$V(xN_0/N)$, the TF energy of the new configuration reads
\begin{equation}\label{scaling}
\U_{\TF}\left(N,V\left(x\frac{N_0}{N}\right), w \right)=
\frac{N}{N_0} \U_{\TF}(N_0,V(x),w),
\end{equation}
while the domain wall contributions remain the same. We can
conveniently consider $N_0$ such that the bulk kinetic energy and
the energy corrections due to the zeros of the density profiles
are negligible with respect to both the TF energy and the domain
wall energy. \emph{A fortiori}, they will be negligible for all
$N>N_0$, since the bulk kinetic energy scales like $N_0/N$.

We consider now an alternative configuration, in which the number
of domain walls is $w'$. The potential energy
$\U_{\TF}(N,V(xN_0/N),w')$ obeys the same scaling law
(\ref{scaling}). If a crossing between the total energies of the
configurations exists, it occurs for a number of particles
\begin{equation}\label{Nstar}
N^*_{w,w'}(\alpha)=N_0
\frac{\Delta\mathcal{E}^{(w)}(\alpha)-\Delta\mathcal{E}^{(w')}(\alpha)}
{\U_{\TF}(N_0,V(x),w')-\U_{\TF}(N_0,V(x),w)},
\end{equation}
which is meaningful only if the differences of the potential
energies and of the domain wall corrections have opposite signs.
Furthermore, physical meaning can be attributed to the crossing
only if $N^*_{w,w'}\geq N_0$, since the validity of TF
approximation is not assured for $N<N_0$.

Binary mixtures of $^{87}$Rb atoms are experimentally available
\cite{hall,tojo}, with mass $m_1=m_2=m=1.45\times 10^{-25}$ kg, in
states $|F = 1,m_F = +1\rangle$ and $|F = 2,m_F = -1\rangle$,
whose $s$-wave scattering lengths are respectively $a_1=5.36$ nm
and $a_2=5.66$ nm. The inter-species scattering length $a_{12}$ is
tunable by approaching a Feshbach resonance \cite{feshbach} (see
\cite{myatt,hall,tojo} for experimental realizations). Let us
suppose that such a mixture is confined in a deformed harmonic
trap, with a longitudinal frequency $\omega_{\ell}=2\pi\times 0.7$
Hz, corresponding to a trapping length
$a_{\ell}=\sqrt{\hbar/(m\omega_{\ell})}= 1.29 \times 10^{-5}$ m
and a transverse frequency $\omega_{\perp}=500\,\omega_{\ell}$,
such that $a_{\perp}=a_{\ell}/(10\sqrt{5})$. Since we want the
transverse degrees of freedom to be freezed, the number of
particles per species $N_0$ has to satisfy $N_0  \ll
a_{\ell}^2/(a_{\perp} \min(a_1,a_2)) \simeq 5\times 10^4$.
Moreover, in order to ensure the applicability of one-dimensional
TF approximation, the condition $N_0 \gg a_{\perp}^2/(a_{\ell}
\max(a_1,a_2)) \simeq 10 $ must hold \cite{stringari}. A good
choice is then $N_0=5\times 10^3$. It is readily verified that if
the potential is (longitudinally) scaled as in (\ref{scaleW}), the
assumption of one-dimensionality and the TF approximation continue
to be valid. The one-dimensional self-interaction parameters read
\begin{equation}
U_{kk}=\frac{2\hbar^2 a_k}{m_k a_{\perp}^2},
\end{equation}
and their ratio $U_{11}/U_{22}=a_1/a_2$ is very close to one.

In order to obtain a double well in the region where the
condensates are trapped, we add to the longitudinal harmonic
potential a cosine potential, so that
\begin{equation}\label{dw}
V(x)=\frac{m\omega_{\ell}^2}{2}x^2+A \cos(Bx),
\end{equation}
with $A/(\hbar\omega_{\ell})=107.75$ and $B a_{\ell}=6.44\times
10^{-16}$.

\begin{figure}
\centering \subfigure[$\,u=180.8$,
$\delta=11304.2$]{\includegraphics[width=0.23\textwidth]{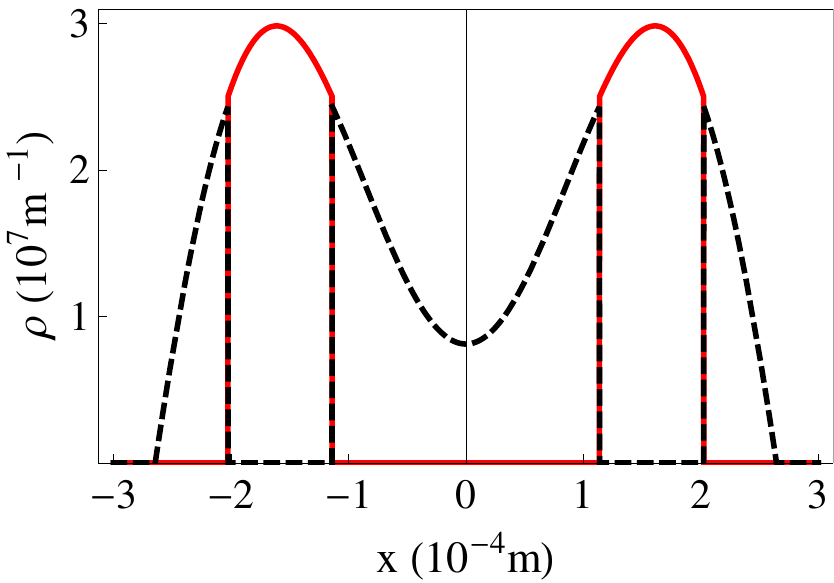}}
\subfigure[$\,u=182.5$,
$\delta=11292.8$]{\includegraphics[width=0.23\textwidth]{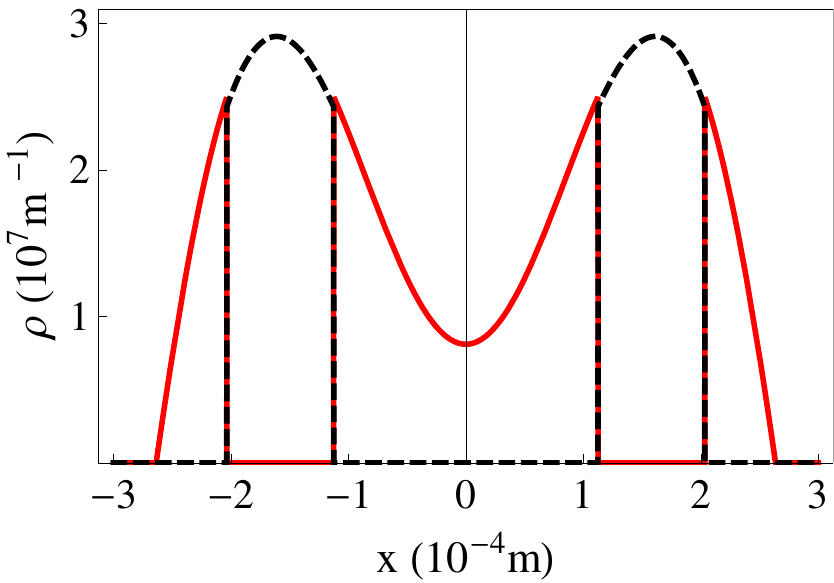}}
\subfigure[$\,u=181.0$,
$\delta=9499.3$]{\includegraphics[width=0.23\textwidth]{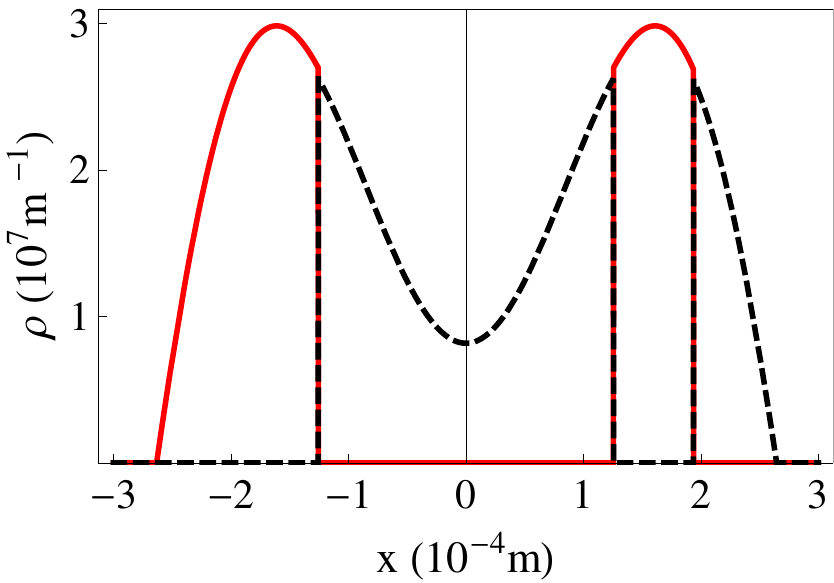}}
\subfigure[$\,u=182.3$,
$\delta=9488.3$]{\includegraphics[width=0.23\textwidth]{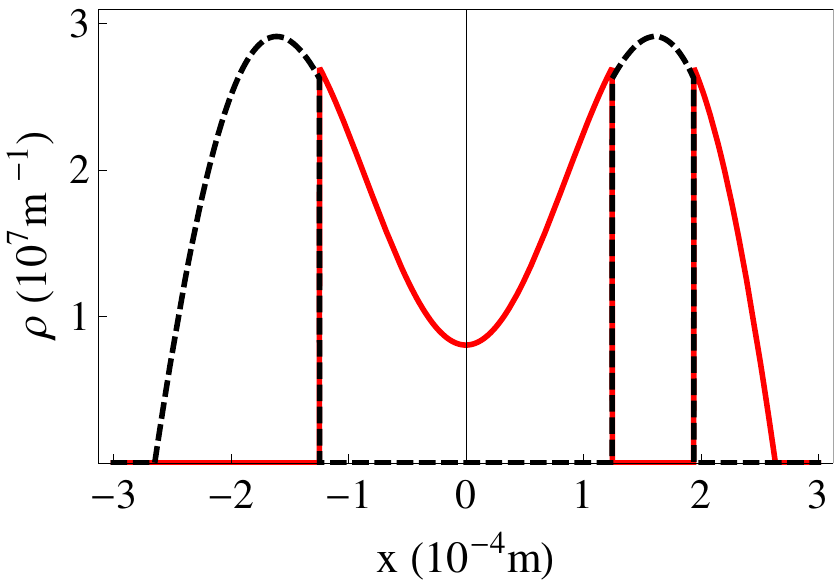}}
\subfigure[$\,u=181.6$,
$\delta=1691.4$]{\includegraphics[width=0.23\textwidth]{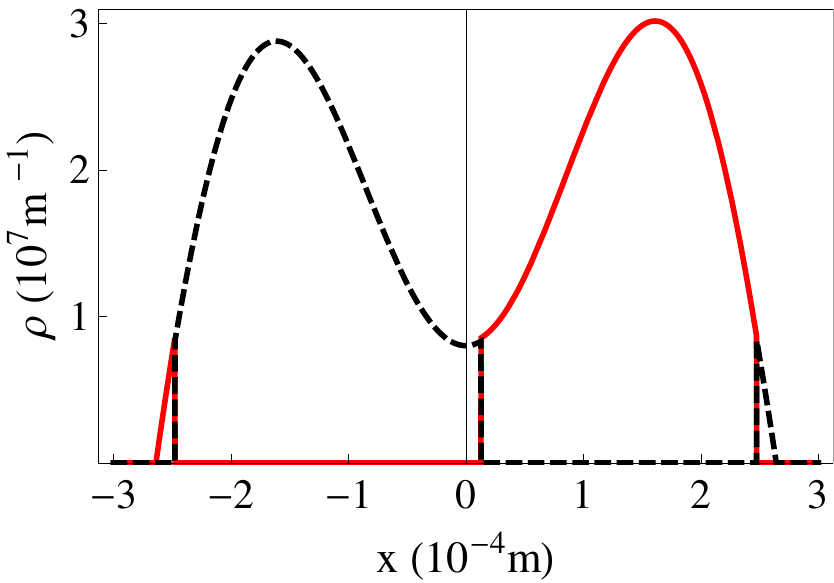}}
\subfigure[$\,u=181.6$,
$\delta=1074.9$]{\includegraphics[width=0.23\textwidth]{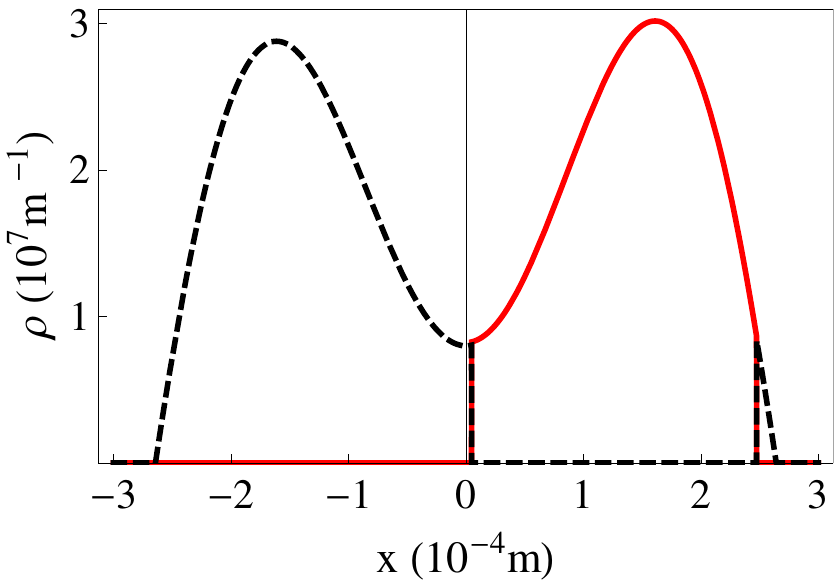}}
\subfigure[$\,u=181.6$,
$\delta=563.8$]{\includegraphics[width=0.23\textwidth]{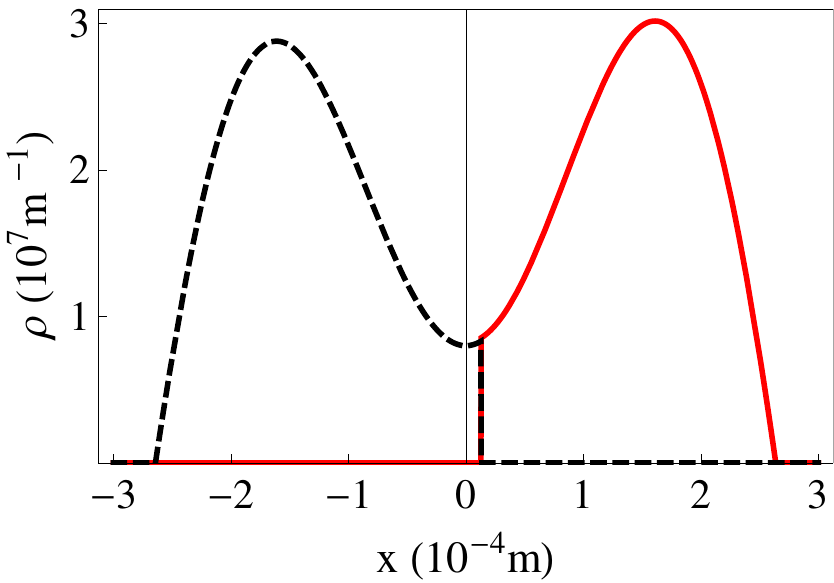}}
\caption{(Color online) Lowest energy configurations in the double
well potential (\ref{dw}), for $N_0=5\times 10^3$ and $\alpha=2$.
In all figures the (linear) densities of species 1 (solid red
line) and species 2 (black dashed line) are plotted vs the linear
coordinate $x$. Below each figure, $u=\U(N_0,V,w)/N_0$ is the TF
potential energy per particle, while $\delta$ is the specific
energy of the domain walls as in Eq. (\ref{walltotal}), both in
units $\hbar\omega_{\ell}$.}\label{fig:profiles}
\end{figure}

\begin{figure}
\centering
\includegraphics[width=0.45\textwidth]{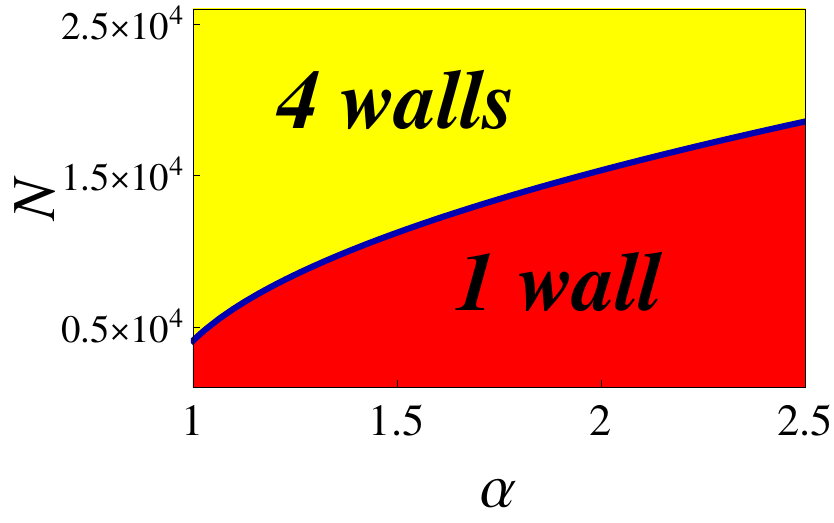}
\caption{(Color online) Ground state phase diagram for a mixture
with equal number of atoms $N_1=N_2=N$ in the potential $V(x)$ of
Eq.\ (\ref{dw}). In the light grey region (yellow in the online
version) configurations with the maximal number of walls are
energetically favored, while in the dark grey region (red in the
online version) the ground state has a single domain wall. The
(blue) transition line represents the function
$N_{1,4}^*(\alpha)$.}\label{fig:diagram}
\end{figure}

The TF stationarity condition (\ref{rhowall}), together with the
normalization conditions, is satisfied by the seven different
configurations represented in Fig.\ \ref{fig:profiles}, together
with their TF and domain wall energies. The number of domain walls
ranges from one to four. The configuration with four domain walls
and with the less-self-interacting species placed in the minima of
the external potential is, as expected, the  minimizer of the TF
energy. However, its domain wall energy is much larger than that
of the configuration with a single domain wall, where each
condensate occupies one well. For instance, $\alpha=2$ yields,
according to definition (\ref{Nstar}), $N^*_{1,4}(2)= 1.4 \times
10^4$. We have observed that for all $\alpha$ the only competing
ground states are the aforementioned configurations, with one and
four domain walls. The four-wall profile is the ground state only
for $N>N^*_{1,4}(\alpha)$, while for smaller $N$ the one with a
single domain wall is energetically favored. In Fig.\
\ref{fig:diagram} the ground state phase diagram is displayed. The
transition line is the graph of the function
\begin{equation}
N^*_{1,4}(\alpha) \propto \Phi_{\alpha}(\bar{\eta}(\alpha))
\end{equation}
and gives direct information about the function
$\Phi_{\alpha}(\bar{\eta}(\alpha))$.  Thus, using Eqs.\
(\ref{Ewall}) and (\ref{Lambda}), the ground state phase diagram
can be used to obtain information about the domain wall energies
and the optimal penetration lengths.

\section{Conclusions and outlook}
\label{sec:conc}

We discussed a variational method that yields a very good
approximation for the total energy of a binary mixture of
Bose-Einstein condensates in a separated configuration. The method
is reliable and accurate. We have seen that, in some cases,
density profiles with a large number of domain walls can be
energetically favored with respect to those with fewer domain
walls, in particular when the interaction ratio $\alpha$ is close
to one. At present, there is a variety of methods to find
approximate solutions to the Gross-Pitaevskii equations, ranging
from analytical techniques \cite{wkb} to numerical ones, including
imaginary-time schemes \cite{qlb,lagrange} and finite-difference
methods \cite{fd1,fd2,fd3}. It is worth emphasizing that our
approach is not aimed at solving the Gross-Pitaevskii equations in
the most general cases, but is rather optimized at uncovering the
ground state properties of the mixture, with minimal numerical
help. This enables us to give an immediate physical interpretation
of the results, and makes it possible to explain in very general
terms the observed phenomena, to predict new ones, and possibly to
develop more refined techniques in order to extend the validity of
our approximations.

All results are analytical and therefore provide solid ground to
improve the approximations. Indeed, starting from Eq.\
(\ref{Ewall}) and from previously obtained TF results
\cite{binbec}, it is possible to compute, for example, corrections
due to the first (finite) derivatives of the potentials, as well
as possible domain-wall displacements and small variations of the
chemical potentials.

The results obtained in this paper can have practical
applications. For example, if one compares an experimental phase
diagram with the theoretical prediction, physical properties of
the mixture can be estimated from the transition line, that
depends on the ratio of the masses and the interaction parameters.
These results can also help in analyzing dynamical phenomena, such
as vortices and solitons, which mostly appear as perturbations of
a stationary background, and lead to very subtle energy changes
(see e.g \cite{coreless,balaz,finl}): in order to correctly
analyze these changes, an accurate estimate of the background
energy is needed.

\end{document}